\documentclass[referee]{raa}
\usepackage{graphicx,times}
\usepackage{natbib}
\usepackage{amssymb,amsmath}
\bibpunct{(}{)}{;}{a}{}{,}

\usepackage[a4paper=true,dvipdfm=true,pagebackref=true]{hyperref}
\hypersetup{pdftitle = The title of my PDF, pdfauthor = My name, pdfsubject= The subject, pdfkeywords = keyword1 keyword2 keyword3}
\hypersetup{colorlinks = true, linkcolor = green, anchorcolor = red, citecolor = blue, filecolor = red, pagecolor = red, urlcolor = red}

\begin{document}

   \title{Three X-ray Flares Near Primary Eclipse of the RS CVn Binary XY UMa
}

 \volnopage{ {\bf 2012} Vol.\ {\bf X} No. {\bf XX}, 000--000}
   \setcounter{page}{1}
\author{Hang Gong\inst{1}, Rachel Osten\inst{2,3}, Thomas Maccarone\inst{4}, Fabio Reale\inst{5,6}, Jifeng Liu\inst{1,7}, Paul A. Heckert\inst{8}}


   \institute{ Key Laboratory of Optical Astronomy, National Astronomical Observatories, Chinese Academy of Sciences, Beijing, 100012, China; {\it ghang.naoc@gmail.com}\\
        \and
             Space Telescope Science Institute, Baltimore, MD 21218, USA;osten@stsci.edu\\
	\and
Johns Hopkins University, Baltimore, MD 21218, USA\\
\and
Department of Physics, Texas Tech University, Lubbock, TX 79409, USA;thomas.maccarone@ttu.edu\\
\and
Dipartimento di Fisica e Chimica, Universit\`a di Palermo, Piazza del Parlamento 1, 90134 Palermo, Italy\\
\and
INAF/Osservatorio Astronomico di Palermo, Piazza del Parlamento 1, 90134 Palermo, Italy\\
\and
College of Astronomy and Space Science, University of Chinese Academy of Sciences, Beijing 100049, China\\
\and
Department of Chemistry and Physics, Western Carolina University, Cullowhee, NC 28723 USA\\
\vs \no
}

\abstract{We report on an archival X-ray observation of the eclipsing RS CVn binary
XY UMa ($\rm P_{orb}\approx$ 0.48d).
 In two $\emph{Chandra}$ ACIS observations spanning 200 ks and almost five orbital periods,
three flares occurred. We find no evidence for eclipses in the X-ray flux.
The flares took place around times of primary eclipse, with
one flare occurring shortly ($<0.125\rm P_{orb}$) after a primary eclipse,
and the other two happening shortly ($<0.05\rm P_{orb}$) before a primary eclipse.
Two flares occurred within roughly one orbital period ($\Delta \phi\approx1.024\rm P_{orb}$) of each other.
We analyze the light curve and spectra of the system, and investigate coronal length scales both during quiescence
and during flares, as well as the timing of the flares.
We explore the possibility that the flares are orbit-induced
by introducing a small orbital eccentricity, which is quite challenging for this close binary.
}

   \authorrunning{Gong\&Osten et al. }            
   \titlerunning{X-ray flares of XY UMa}  
   \maketitle

%
\section{Introduction}           
\label{sect:intro}

X-ray studies of short orbital period systems provide the opportunity to investigate coronal structures
by investigating the phase dependence of emission, with the advantage that multiple orbital periods
are often accessible.
Due to the effect of tidal locking and the increase of stellar magnetic activity
with decreasing rotation period \citep{2014ApJ...794..144R}, some short orbital period systems
display enhanced magnetic activity. However, the phenomenon of supersaturation of the X-ray emission can also occur,
leading to a decrease in the level of observed magnetic activity \citep{2011ApJ...743...48W}.
Eclipsing systems additionally allow for constraints on the extent of
X-ray emitting material above the stellar photosphere.
Previous X-ray studies of short-period systems have found a lack of strong X-ray eclipses, with
suggestions of high latitude, compact coronae, from systems as short period as 0.27 d (contact binary systems
VW~Ceph and 44 iBoo; Huenemoerder et al. 2006; Brickhouse et al. 2001 respectively ).
A recent study of the M dwarf eclipsing binary YY~Gem \citep{2012MNRAS.423..493H}, a P$_{\rm rot}=P_{\rm orb}=$0.81 d, M1V+M1V binary,
found that there were no strong X-ray eclipses, and both components were active.
Other studies have suggested a causal connection between the timing of flares and periastron passage in
close binaries \citep{2002A&A...382..152M,2008A&A...480..489M,2011ApJ...730....6G}, with the interpretation that of
interacting magnetospheres of two systems.
These stellar systems can also provide a context in which to place star-planet interactions with close-in,
magnetized exoplanets \citep{2000ApJ...529.1031R}.

RS Canum Venaticorum systems (RS CVns hereafter, Hall 1976, 1989) are close but detached binaries, typically with a G/K giant or subgiant + a
late-type main sequence/subgiant companion. Regular RS CVns have orbital periods between 1 and 14 days, while systems with short periods less
than one day can also exist. Tidal locking enhances chromospheric and coronal emission, making RS CVns among the most magnetically active
late type stellar systems (see the introduction of Osten \& Brown 1999). Since its discovery in 1955, XY UMa has been one of the most
intensively observed RS CVn binaries. It has the fifth shortest orbital period ($\approx$0.47899d) according to the RS CVn binary
catalog \citep{2005A&A...437..375D}, which means both of its two companions, G2-3V+K4-5V \citep{1993A&AS..100..173S,1997AcA....47..451P},
should have become tidally locked \citep{1977A&A....57..383Z,1988ApJ...324L..71T,1992ApJ...395..259T,2006ApJ...651.1151A,2006ApJ...653..621M,2008EAS....29....1M}.
As described in \citet{2001A&A...371..997P}, there are two different distances, one based on Hipparcos astrometric data (66$\pm$6 pc)
and the other based on the absolute magnitudes of the two companions (86$\pm$5 pc).
Here, we adopt the second value because of Hipparcos' short-term coverage and the possible perturbation on the astrometry by a third object.
The binary parameters are
$\rm{R_{1}}$=1.16 $\rm R_{\odot}$, $\rm{R_{2}}$=0.63 $\rm R_{\odot}$, the semi-major axis a=3.107 $\rm R_{\odot}$ and
the orbital inclination i=80.86$^{\circ}$, while
\citet{1998A&AS..127..257E} derive the orbital inclination of 76$^{\circ}$, relative to an edge-on system of 90$^{\circ}$.

\citet{1994MNRAS.267.1081H} showed XY UMa had substantial star spot activity based on 1189 V-band photometric observations in 1992 October.
They claimed the existence of a dark zone encircled the primary star between $\pm15^\circ$ in latitude and spot accumulation at the inner hemisphere
of the primary. Also based on substantial optical photometric data, \citet{1997MNRAS.287..567C} and \citet{2001MNRAS.326.1489L} used eclipse
mapping to map the spot distribution. Their results are consistent with \citet{1994MNRAS.267.1081H} and showed the spot evolution in few days to
one week. \citet{1995MNRAS.277..747H} interpreted long-term photometric variations of XY~UMa as originating partly in a polar spot
on the primary, which dominates the optical photometric variability.

Although XY UMa is magnetically active and has long term photometric observations in the past several decades, relatively few optical flares
have been seen \citep{1990MNRAS.246..337J}. Only two flare-like events were reported in optical band. \citet{1983AJ.....88..532Z} discovered
one between phases 0.54 and 0.62 at UBV bands, while the other \citep{2001A&A...366..202O}, derived from the excess emission in $\rm H_{\alpha}$,
occurred between phases 0.6 and 0.8. The lack of optical flare detections could be result of flare-induced brightenings being relatively
small compared with XY UMa's overall optical brightness.
XY UMa is also X-ray bright. It was observed by EXOSAT for a continuous 14.5 hours in 1986. A moderately higher count rate between phases 1.41 and 1.47 for about 1.5 ks was interpreted as a flare event by \citet{1990MNRAS.246..337J}, but it was not mentioned by a previous analysis \citep{1990MNRAS.243..557B} of the same data. XY UMa was also observed by ROSAT for a total 37 ks in 1992. An enhanced count rate was detected at phase 0.5 exactly in the folded light curve of different observations \citep{1998MNRAS.295..825J}. In general,
three flares out of the four noted in the literature occur near the secondary eclipse.

The paper is organized as follows: \S 2 describes the observations and data reduction; \S 3 describes what we can derive about the
coronal length scales in both quiescence and during flares, as well as what we can say about the timing of the X-ray flares.
\S 4 has a discussion of the results and \S 5 concludes.

\section{Chandra Observation And Data Reduction}
As Table.1 shows, XY UMa was observed by $\emph{Chandra}$ twice, with observations separated by five days in 2001 April. The two observations were similarly configured except for the exposure time.
The original purpose of the observations was an X-ray cluster survey.
In processing all of the publicly available ACIS data while searching for flare-like events, we
discovered three flares from XY UMa.

The large off-axis angle of XY UMa introduces several technical problems.
One concern is whether XY UMa has dithered off the detector\footnotemark[1]. The standard dither pattern of the $\emph{Chandra}$ telescope
is 16$\arcsec$, but the position of XY UMa is about 1$\arcmin$ away from the edge of the detector. The large off-axis angle
combined with the spokes (Figure 4.14 of $\emph{Chandra}$'s POG 18\footnotemark[2]) in the image may attenuate the count rate of XY UMa,
but the profile of the light curve can be recovered. We also assess whether XY UMa is affected by pileup\footnotemark[3] because of
its brightness. The script $pileup\_map$\footnotemark[4] of CIAO 4.7 returns 0.067 (count/event island/frame time)
near the centroid of XY UMa(ObsID=2227), which corresponds to $<$ 5\% pileup fraction in the 3x3 pixel cell of the most brightest region,
where there are only about 3000 counts.

\begin{table}
\caption{$\emph{Chandra}$ Observations of XY UMa}
\begin{center}
\begin{tabular}{lllllll}
\hline\hline
Instrument & Date & ID  &Exposure Time&Counts$^{a}$   & Off-axis angle &Size$^{b}$ \\
     \hline
ACIS-I&2001 Apr 24&2452&76 ks(1.84$\rm P_{orb}$)&19579/18295&11.23$\arcmin$&12.1$\arcsec$x10.0$\arcsec$\\
\hline
ACIS-I&2001 Apr 29&2227&124 ks(3.00$\rm P_{orb}$)&46302/43302&11.32$\arcmin$&12.0$\arcsec$x10.0$\arcsec$\\
\hline
\end{tabular}

\end{center}
\footnotesize{$^a$}{0.3-10 keV photons in the 3$\sigma$ elliptical region by $wavdetect$ and a 10$\arcsec$ radius circle respectively}\\
\footnotesize{$^b$}{axis length of the 3$\sigma$ elliptical region}\\
\end{table}

\subsection{The Phase}
Because our analysis makes use of the times of primary and secondary eclipse during the
X-ray observations, we use an optical light curve from \citet{2012JAD....18....5H}.
This is the optical observation closest to the time of the $\emph{Chandra}$ observations we can find,
taken just 20 days after the $\emph{Chandra}$ observations.
\citet{2001MNRAS.326.1489L} showed the zero point of the ephemeris in 2000 has about half an orbital period shift compared with the zero point in 1999.
This means it is necessary to adopt an observation close in time to keep the phase coherence.
Hence, the phase uncertainties here are not from the ephemeris \citep{2001MNRAS.326.1489L}, but mainly from the
orbital period ($< 1\times10^{-5}$ d, Chochol et al. 1998; Pribulla et al. 2001) and the timing precision of every V
band photometric point ($\approx 1\times10^{-4}$ d) given by \citet{2012JAD....18....5H}. The combined temporal
uncertainty of each point in the V band light curve is less than 10 minutes.

\subsection{X-ray Light Curves}
Data reduction began with reprocessing the level 2 data for
both observations by $chandra\_repro$ to ensure consistent calibration updates and the newest software are applied.
Then, X-ray photons between 0.3-10 keV were extracted from the $3\sigma$ elliptical region derived by
$wavdetect$\footnotemark[5], the workhorse of CIAO for source detection. If we use a 10$\arcsec$ radius circle as the source region, the X-ray counts do not vary much. Because the optical data described in \S 2.1 is in HJD, the X-ray timing data are also converted to Heliocentric Julian Day (HJD) according to web pages\footnotemark[6]. The HJD correction is
no more than 1 minute compared with Julian date. We adjust the time bin scale and find a 200 s time binning reduces uncertainties in each light curve bin while retaining
information about any temporal variations.

Figure.1 shows the X-ray light curves, along with the hardness ratio, using the 200 s binning.
The hardness ratio (HD ratio) is defined as (h-s)/(h+s), in which $s$ is the number of counts in a soft band (0.3 to 2 keV),
while $h$ is the number of counts in a hard band extending from 2 to 10 keV.
A roughly synchronous evolution can be seen compared with X-ray light curves.
Figure.1 also overplots the V band photometry, to illustrate the times of primary and secondary eclipse.
As Figure.1 shows, two large flares (f1 and f2 hereafter) occurred shortly before two primary eclipses in the
second observation, while a smaller flare (f3 hereafter) occurred in the first observation five days earlier. The peak
count rates of f1, f2 and f3 increase with a factor of about 4.2, 2.6 and 2.4 respectively compared with the base level {(between 50 counts/bin and 75 counts/bin} of the light curve selected by eye.
We note key times in the light curve: b1 and s1 are the approximate beginning and ending, respectively, of f1, b2 and
s2 are the corresponding quantities for f2. The peaks of f1 and f2 are denoted p1 and p2, respectively, and
are the locations of local maxima in the light curves of f2 and f2.
The location of optical primary eclipses are termed e0, e1, and e2, and are also marked in the light curve.

Because of the asymmetrical profiles and the relatively sparse counts near the flare peaks, we directly adopt points with the biggest counts, p1 and p2, as flare peaks.
The separation between the peak of f1 and its corresponding primary eclipse is about 71.2 minutes, and for f2, it is about 54.2 minutes. The difference (17min $\approx 0.024\rm P_{orb}$) could be bigger or smaller depending
on where the flare peaks and where primary eclipses are exactly.
Using a Possion error for every time bin and the method of
least squares, assuming an exponential $y=a \times e^{-t/\tau}+b$ decay, we obtain an e-folding decay time $\tau$=1433$\pm$249s
 ($\chi^{2}$=1.9, dof=31) for [p1, s1], while $\tau$=3732$\pm$1236s ($\chi^{2}$=1.4, dof=26) for [p2, s2]. The decay continues between s1 and b2,
but the exponential fit is poor because of fluctuations after s1.

\begin{figure}
   \centering
   \includegraphics[width=14.0cm, angle=0]{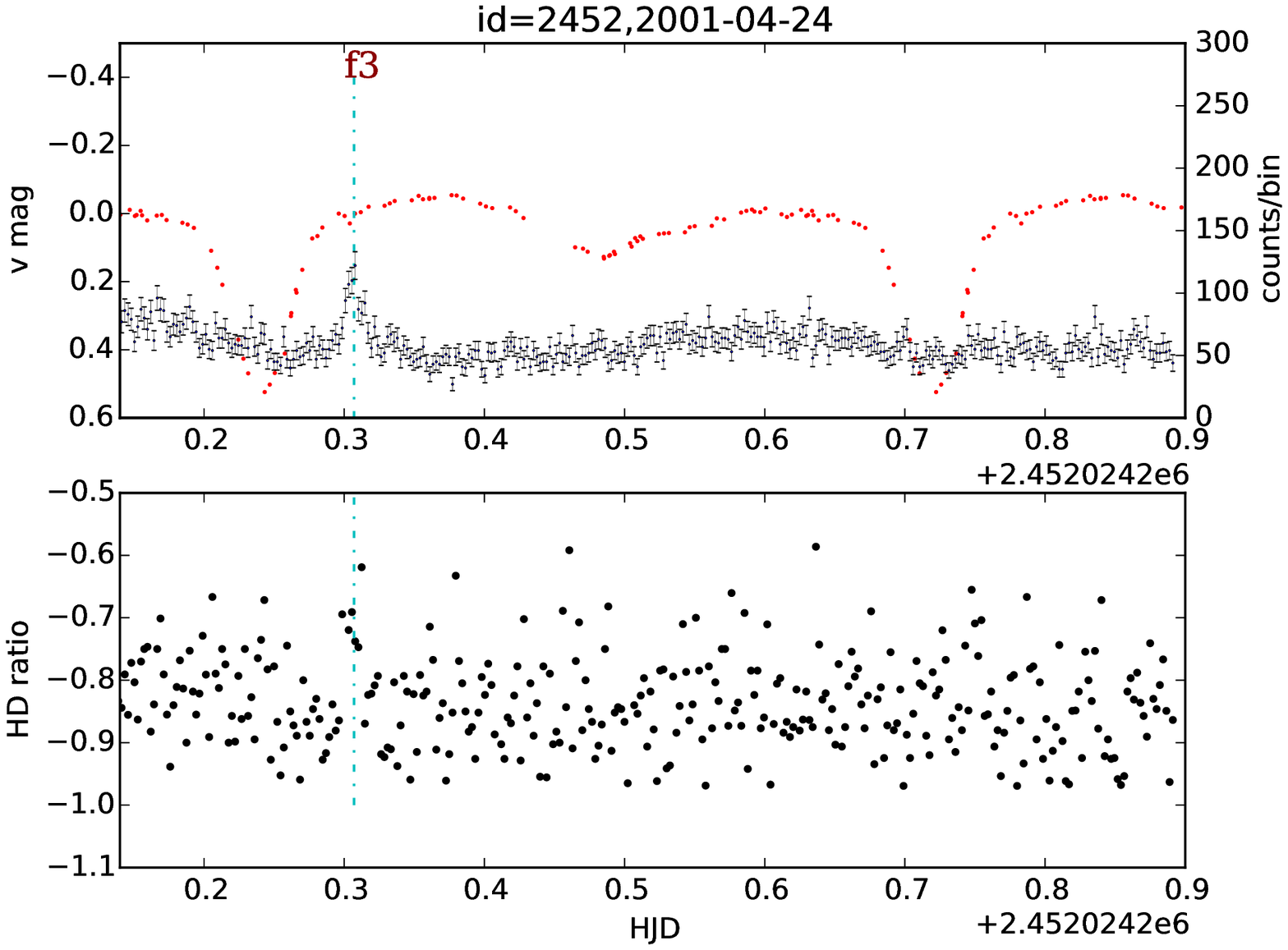}
   \includegraphics[width=14.0cm, angle=0]{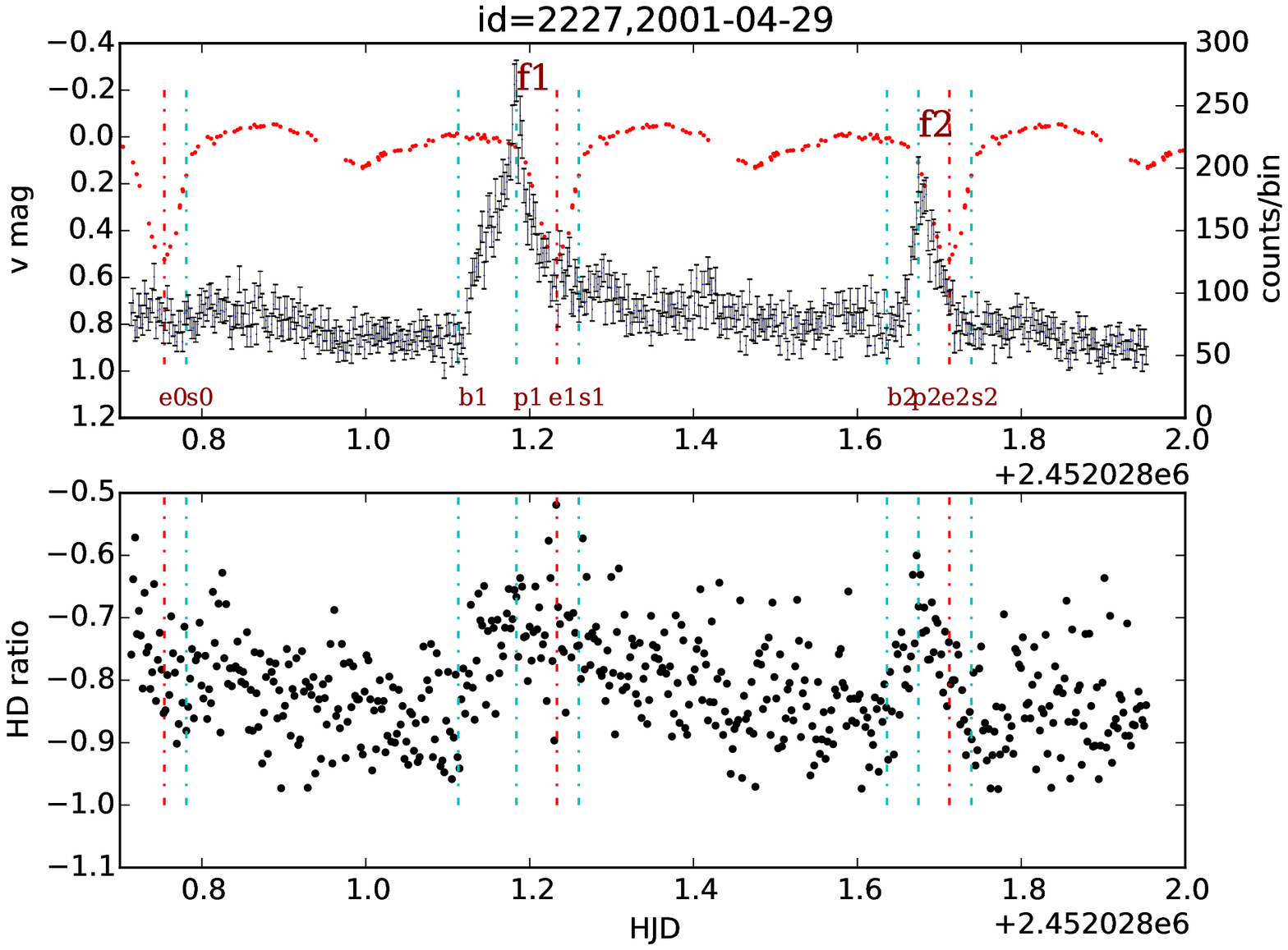}
   \caption{X-ray light curves from the two $\emph{Chandra}$ observations}
{\footnotesize In the two upper panels, the blue points are from the X-ray data, while the red points are from real optical observations which have a 20 day offset. The hardness ratios are in the lower panels. Critical timing points selected by eye and primary eclipses are marked by cyan and red dashed lines respectively in all panels.}
   \end{figure}

\subsection{X-ray Spectra}
The X-ray spectrum of coronally active stars is well-described by a collisionally ionized plasma, and we
use the absorbed ($xswabs$\footnotemark[7]) APEC model \citep{2001ApJ...556L..91S} implementation in Sherpa to fit each spectrum.
Time-resolved X-ray spectra were extracted using the light curve time intervals noted in
Figure.1. The [p1,s1] and [s1,b2] temporal intervals were subdivided into two and three equal parts, respectively, to get some resolution,
and spectra were extracted from each sub-interval. Source and background spectra were created by $specextract$ and fit by an absorbed two-temperature APEC model.
We freeze the redshift in APEC to be zero, leave all other parameters free, and use the Sherpa $moncar$ (the Monte Carlo optimization method) to search for the best model parameters.

The value of N$_{H}$ is not constrained, but N$_{H}$ obtained are all near zero.
We settled on using two temperature components to describe the spectra: a one-temperature APEC model produced
large $\chi^{2}$ values, indicating an inadequacy of the spectral fits. Spectra corresponding to non-flare
temporal bins could also be described by a four-temperature APEC model, but the improvement is not enough.
The non-flare parts are relatively poorly fit.
Results are shown in Table.2.

The spectra of f1 ([b1,s1]) and f2 ([b2,s2]) are shown in Figure.2. Both have a 2.1 keV instrumental absorption edge (Table 9.4 of $\emph{Chandra}$'s POG 18; Moran et al. 2005), which also indicates photon pileup can be ignored.
Taking the integrated flux from the segments within [b1,s1] and [b2,s2] in Table.2 and
multiplying by the integration times and assumed distance of 86 pc, the radiated energies in the
0.3-10 keV bandpass for f1 and f2 are
$1.1\times10^{35}$ erg and $5.5\times10^{34}$ erg, respectively.

\begin{figure}
   \centering
   \includegraphics[width=14.0cm, angle=0]{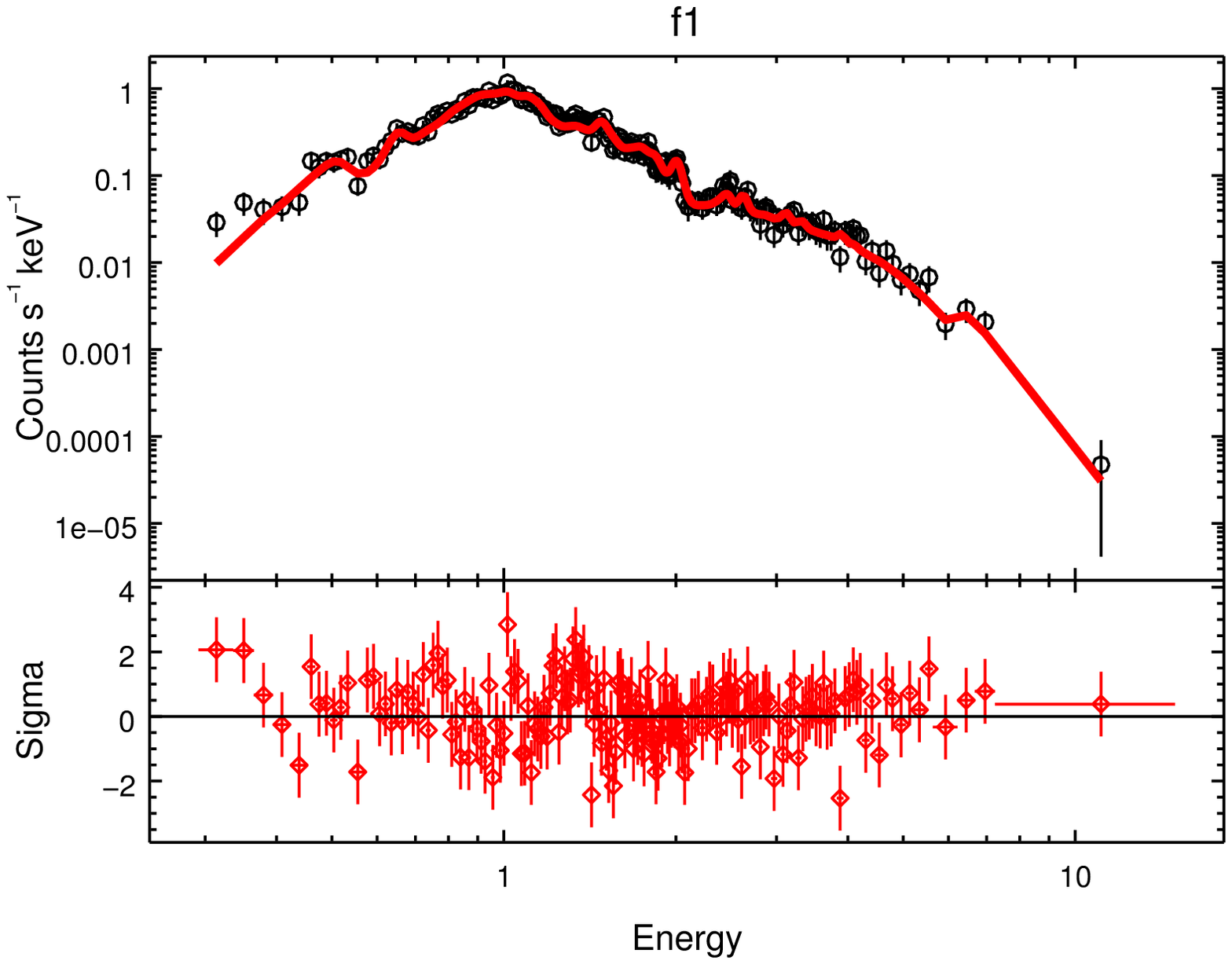}
   \includegraphics[width=14.0cm, angle=0]{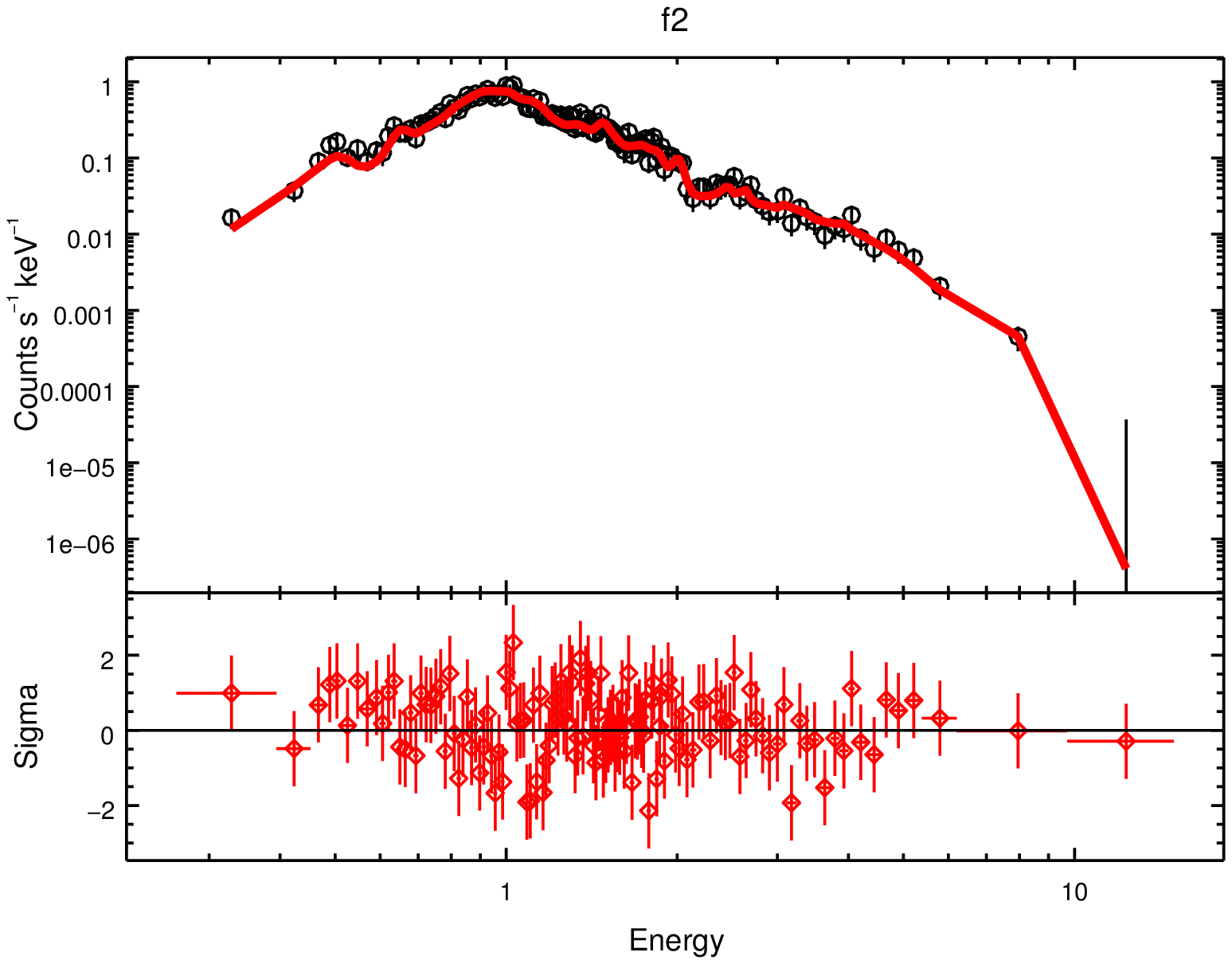}
   \caption{An absorbed-2T fit for f1 and f2. The high temperature part evolves, while the low temperature part sustains at about 1 keV.}
   \end{figure}

\section{Analysis}

\subsection{Quiescent Coronal Length Scales}
Despite the fact that the X-ray observations span multiple binary eclipses, Figure.1 reveals that
there is no apparent diminution of the X-ray flux
during these events. The Volume Emission Measures (VEMs) reported in Table.2 outside of flares
provide a constraint on emitting volumes and hence length scales, for assumptions about the coronal electron density.
The volume emission measure in Table.2 is $VEM=n_{e} n_{H} dV$, with $n_{e}$ the coronal electron density,
$n_{H}$ the number density of Hydrogen, and $dV$ the emitting volume. For a fully ionized plasma, $n_{H}/n_{e}=$0.8.
The interval between 0 and b1 in Table.2 should occur during a primary eclipse, according to
Figure.1, and we use the emission measures reported to investigate physical length scales.
\citet{2004A&A...427..667N} examined coronal electron densities derived from density-sensitive X-ray line
diagnostics, for a range of stellar activity levels. Densities derived from the \ion{Ne}{9}
triplet, formed at a temperature around 4 MK, ranged from $\log n_{e}$ (cm$^{-3}$)=10.5$-$12, and they found
no conclusive trend of densities with activity level for the higher activity stars.
As this temperature is closest to the lower of the two temperatures returned from our spectral fitting,
we use the emission measure reported in Table.2 for the 0.92 keV plasma (0.92 keV$\approx$11 MK),
which is $n_{e} n_{H} dV=$5.27$\times$10$^{27}$ cm$^{-3}$.  Evaluating at the low and high ends of the
electron densities found in \citet{2004A&A...427..667N}, the coronal volume ranges between 6.6$\times$10$^{28}$ cm$^{3}$
and 6.6$\times$10$^{31}$ cm$^{3}$. This is consistent with the volume trends described in \citet{2003ApJ...582.1073O}.

If this volume is distributed homogeneously over the surface of one star, then the height of the coronal shell can be estimated from \\
\begin{equation}
V_{\rm cor} = 4 \pi R_{\star}^{2} h_{\rm cor}
\end{equation}
where $V_{\rm cor}$ is the coronal emitting volume, R$_{\star}$ is the stellar radius, and $h_{\rm cor}$ is
the height of a spherically symmetric X-ray-emitting region. Evaluating separately for each star, we find\\
\begin{eqnarray}
h_{\rm cor,1}& =&8\times10^{5}-8\times10^{8} \;\;\; cm  \\
h_{\rm cor,2}& =&2.7\times10^{6}-2.7\times10^{9} \;\;\; cm
\end{eqnarray}
for the primary and secondary, respectively, with $R_{1}= 1.16 R_{\odot}$ for the primary and $R_{2}=0.63R_{\odot}$
for the secondary.
These sizes are very small compared to the binary separation (3.1 R$_{\odot}$ = 2.2$\times$10$^{11}$ cm),
and would suggest that the X-ray-emitting material would be eclipsed as well. In order for this not to be
the case, the coronal material should be at a high latitude which is always visible.

The geometry of the system provides another constraint on the coronal length scales during quiescence.
Assuming that the orbital and rotational axes are aligned, the constraint on the orbital inclination
of $\i=14^{\circ}$ from \citet{1998A&AS..127..257E} suggests that an emitting region on the surface of either star
would need to be located between $0$ and $14$ $^{\circ}$ co-latitude (i.e., at the visible pole or up to
14$^{\circ}$ away in latitude).  For an extended structure, we use the formalism described in
\citet{1994ApJ...430..332L}, using the quantity $\xi$, \\
\begin{equation}
\xi = \sin \theta \sin i \cos \phi + \cos \theta \cos i
\end{equation}

where $\theta$ is the co-latitude of the emitting region, $\phi$ is the longitude, and $\i$ is the inclination (with a
90$^{\circ}$ offset in the orbital inclination used above).
An emitting region above the surface is visible as long as $\xi > 0$ and the relation\\
\begin{equation}
\left( 1+ \frac{h}{R_{\star}} \right) \sqrt{(1-\xi^2)} > 1
\end{equation}
holds. We evaluated these conditions for each star to determine the minimum height required to be visible at all longitudes,
and found that a relative height of 0.03 R$_{\star}$ was sufficient. For the primary, this works out to $\approx$2.5$\times$10$^{9}$
cm, and for the secondary it is 1.4$\times$10$^{9}$ cm.

\subsection{Flaring Length Scale}
The light curve and spectral evolution of flaring plasma hold information about the flaring coronal length
scales, once assumptions about the release of energy and geometry are made.
Most often, analyses of the decay phase of stellar coronal flares are used to reveal a loop
semi-length using the method of \citet{1997A&A...325..782R}.
\citet{2007A&A...471..271R} gives an empirical relation based on max(temperature)
$\rm{T_{0}}$, max(emission measure), the time ($t_{M,3} $) at which max(EM) occurs and the temperature $\rm{T_{M}}$ when max(EM) occurs.
In light of the poor statistics given by our spectral fitting of sub-intervals of the decay phase of f1,
we use this alternate method to estimate the size of a single flaring loop.
We evenly divide the decay phase of [p1,s1] and [s1,b2] respectively, and have five temperature bins.
As Table.2 shows, emission measure drops
sharply after [p1,s1]a (@9.42 ks), while the temperature sustains around 3.2 keV.
Using Equation(12) of \citet{2007A&A...471..271R}, \\
\begin{equation}
\rm{L_{9}}
\approx 3 {\Psi}^2 \sqrt{\rm{T_{0,7}}}  t_{M,3}
\end{equation}
where $\rm{T_{0}}$ is the maximum temperature attained during the flare, $\rm{T_{0,7}}$ is $\rm{T_{0}}$ in units of 10$^{7}$K, $\rm{t_{M}}$
is the time at which the maximum emission measure occurs, $t_{M,3}$ is $\rm{t_{M}}$ in units of 1000 seconds, and  $\Psi$ is $\frac{\rm {T_{0}}}{\rm {T_{M}}}$.
Evaluating this for f1,  we let $\rm{T_{0}}\approx T_{M}$=3.2 keV and $t_{M,3}$=(102+55)minutes=9420s, and get the
loop half length $\rm{L}\approx 3\times {(\frac{3.2}{3.2})}^2 \times \sqrt{3.2\times1.16}\times9.42\times 10^{9}$ cm=0.78 $\rm R_{\odot}$.
Then the loop height is 0.50 $\rm R_{\odot}$ assuming a vertical and circular loop.
Hence, if it is one
single loop, it is not long enough to anchor on the two companions simultaneously, but the loop height is
a significant fraction of the separation between
the two companions, and thus there could be magnetosphere interaction in between.

\begin{table}
\caption{APEC Fit \label{tbl:specfit}}
\begin{tabular}{ccccccccc}
\hline\hline
Phase&Duration&APEC.2T&$\chi^{2}$(dof)&Flux$^{a}$&Luminosity$^{b}$&EM\\
.........&(minute)&(keV)&.........&...&($\rm{10^{30}\ erg\ s^{-1}}$)&($\rm{10^{53}}cm^{-3}$,$\rm{10^{52}\,cm^{-3}}$)\\
\hline
[0, b1]&576&$2.62^{+0.29}_{-0.09},0.92^{+0.01}_{-0.01}$&2.4(171)&4.4&3.9&$1.42^{+0.04}_{-0.04},5.27^{+0.21}_{-0.21}$\\
\hline
[b1, p1]&102&$2.98^{+0.18}_{-0.18},0.97^{+0.03}_{-0.03}$&0.9(127)&9.7&8.6&$3.83^{+0.14}_{-0.14},7.17^{+0.60}_{-0.61}$\\
\hline
[p1, s1]a&55&$3.20^{+0.26}_{-0.24},1.00^{+0.03}_{-0.03}$&0.9(102)&11.1&9.8&$4.25^{+0.21}_{-0.21},8.77^{+0.90}_{-0.90}$\\
\hline
[p1, s1]b&55&$3.22^{+0.33}_{-0.30},1.00^{+0.03}_{-0.04}$&0.7(78)&7.4&6.5&$2.80^{+0.18}_{-0.18},6.20^{+0.79}_{-0.79}$\\
\hline
[s1, b2]a&180&$2.87^{+0.18}_{-0.17},0.98^{+0.02}_{-0.02}$&1.4(129)&5.8&5.1&$2.11^{+0.08}_{-0.08},5.61^{+0.37}_{-0.37}$\\
\hline
[s1, b2]b&180&$2.24^{+0.14}_{-0.12},0.92^{+0.03}_{-0.03}$&1.1(116)&4.9&4.3&$1.69^{+0.08}_{-0.08},5.43^{+0.42}_{-0.46}$\\
\hline
[s1, b2]c&180&$2.52^{+0.18}_{-0.18},0.94^{+0.03}_{-0.03}$&1.5(106)&4.4&3.9&$1.34^{+0.07}_{-0.07},5.64^{+0.40}_{-0.40}$\\
\hline
[b2, p2]&55&$3.57^{+0.54}_{-0.39},0.96^{+0.03}_{-0.04}$&0.8(72)&6.4&5.6&$2.13^{+0.15}_{-0.16},6.39^{+0.69}_{-0.69}$\\
\hline
[p2, s2]&93&$3.22^{+0.26}_{-0.24},0.97^{+0.02}_{-0.03}$&0.9(106)&7.3&6.4&$2.54^{+0.13}_{-0.13},7.17^{+0.60}_{-0.59}$\\
\hline
[s2, end]&309&$2.57^{+0.16}_{-0.15},0.93^{+0.02}_{-0.02}$&1.8(126)&3.7&3.3&$1.04^{+0.05}_{-0.05},5.30^{+0.28}_{-0.28}$\\
\hline
f1:[b1, s1]&212&$3.09^{+0.12}_{-0.12},0.98^{+0.02}_{-0.02}$&1.1(173)&9.7&8.6&$3.79^{+0.09}_{-0.09},7.47^{+0.41}_{-0.41}$\\
\hline
f2:[b2, s2]&148&$3.31^{+0.22}_{-0.19},0.97^{+0.02}_{-0.02}$&0.9(129)&7.0&6.2&$2.41^{+0.10}_{-0.10},7.00^{+0.44}_{-0.44}$\\
\hline
\end{tabular}

\footnotesize{$^a$}{absorbed flux between 0.3-10 keV in unit of $10^{-12}\rm{erg\, cm^{-2}\, s^{-1}}$}\\
\footnotesize{$^b$}{luminosity based on absorbed flux}\\

\end{table}

\subsection{Timing of the flares}

RS CVns should flare more frequently than typical binaries in the X-ray band, due to their shorter
orbital and hence rotational periods.
However, it is difficult to assess how frequently a particular RS CVn binary flares since
long term X-ray observations of any one system are lacking. \citet{1999ApJ...515..746O} analyzed 12.2 Ms EUVE
photometric data of 16 RS CVn binaries and partly answered this question. Of the dozens of flares,
only a few had peak flare count rates increase by more than a factor of three compared to the non-flaring
count rates.  As noted in \S 2.2, the peaks of these flares are factors of 4.2, 2.6, and 2.4 above a non-flaring
count rate, and the integrated energies derived in \S 2.3 also reveal these to be large releases of energy.

The energetic releases of these two
flares are fairly large and should occur relatively rarely.
Here we attempt to quantify this by extrapolating from what is known about
flares on RS~CVn systems as well as single stars.
\citet{2000ApJ...541..396A} characterized the coronal flare frequency of active single G and K dwarfs
as a function of the star's X-ray luminosity, \\
\begin{equation}
N(>10^{32} erg) = 1.9\times10^{-27} L_{x}^{0.95} \rm{\ flares/day}
\end{equation}
above a flare energy of 10$^{32}$ erg,
and we use this to estimate the number of flares which would be expected to occur on
one of the stars in the XY~UMa system,
assuming that the flare frequency distribution for tidally locked binary systems is
similar to that of single active stars.
We compute $L_{x}$ using the values for quiescence in Table.2,
and dividing the observed X-ray luminosity (3.9$\times$10$^{30}$ erg s$^{-1}$) equally between the two stars.
\citet{1999ApJ...515..746O} investigated the flare frequency versus
energy distribution for the 16 RS~CVn systems mentioned above, and characterized it by an index $\alpha$ near 1.6,
where the differential number of flares occurs per unit time per unit energy as $dN/dE \propto E^{-\alpha}$.
More recent investigations of flare frequency distributions for active stars have revealed a range of $\alpha$
going up to about 2.2 \citep{2007LRSP....4....3G}, and we consider this range here, as the precise flare frequency distribution for
XY~UMa is not known. Given the flare rate in \citet{2000ApJ...541..396A}, the number of flares expected a critical level
$E_{\rm crit}$ is, \\
\begin{equation}
N_{\rm expected} = N_{\rm tot} \left( \frac{E_{\rm crit}}{E_{\rm min}} \right)^{1-\alpha}
\end{equation}
with $E_{\rm min}$=10$^{32}$ erg, and $N_{\rm tot}=N(>10^{32} erg)\times \Delta t$ flares.
For $\alpha=$1.6,1.8,2.0,2.2, and a critical flare energy of $1.1\times10^{35}$ erg (the radiated energy
of the f1 flare), we would expect
2.4,0.6,0.1,0.04 flares in the 1.44 days of the 29 April observation. Thus these flares are consistent or marginally consistent with
the expected number of flares for the flatter distributions, but for the steeper distributions
($\alpha \geq$ 2) they are incompatible. At this point the large loop lengths coupled with particular
timing of flares relative to primary eclipse are suggestive but not conclusive of magnetospheric interaction
in this binary system.

\section{Discussion}
It is curious that two of the flares observed on XY UMa occurred within 0.05 P$_{\rm orb}$ of primary eclipses of the binary system. We have investigated
via flare loop hydrodynamic modelling whether the flaring structures are large
enough to enable interactions between the two stars in the binary system, given
how close they are. The separation of the two stellar surfaces is only
a few stellar radii,
making it plausible that a flaring loop of height 0.5 $\rm R_{\odot}$ could
possibly interact with a similarly sized loop on the other star of the binary system.
The fact that the X-ray observations do not show any evidence
of primary or secondary eclipses indicates fairly extended coronal structures. The flares themselves
exhibit a fairly classic rise and decay light curve structure without evidence for
eclipses of the flaring material.

The similar phases of the three sporadic flares near primary eclipse make XY UMa a stunted version of CF Tuc \citep{1997MNRAS.287..199G}. Also as an eclipsing RS CVn system, with a 2.78d orbit, CF Tuc shows a clear modulation with a radio-flux maximum at phase 0.5, which is caused by an active intra-binary region probably. V711 Tau shows a similar behaviour seen here, in that two nearly identical flares,
which both increase by a factor of about two, are separated by $\approx \frac{2}{3}\rm P_{orb}$.
The binary $\sigma^{2}$ CrB also displayed flares separated in phase by nearly two orbital periods
\citep{2000ApJ...544..953O}. It is also interesting to find some potential mechanism responsible for the phases of f1, f2 and f3,
especially when the phases of f1 and f2 are almost the same.

The timing of flares on XY UMa, both those studied here and reported elsewhere in the
literature, appear to occur preferentially near primary or secondary eclipse.
Previous studies of activity on RS~CVn systems have shown a clustering of starspots at
preferential, or ``active'' longitudes \citep{2006Ap&SS.304..145O}.
Since coronal flares are presumed to originate from the regions of concentrated
magnetic field which manifest in the photosphere as starspots or active regions,
the existence of two flares separated by nearly an orbital period suggests that an active region
at the same longitude could be the origin for both events.

Another possibility to explain the particular timing of the flares is that
there is some mechanism to trigger flares near a primary eclipse.
Periastron-induced activities such as those described in \citep{2002A&A...382..152M,2008A&A...480..489M},
could cause the interaction of the two magnetospheres or a joint-magnetosphere, and thus
trigger magnetic reconnection flares during close approach.
This scenario would require XY UMa's periastron to be near its primary eclipse.
Based on the loop length analysis in \S 3.2, we
find weak evidence for the f1 flare to originate in an extended structure.

It is quite challenging (see the six reference papers in our introduction) to have an eccentric orbit for XY UMa. The question is how circular its orbit is or how small its orbit eccentricity is. Three factors lead us to reconsider its supposed circular orbit. First, a tertiary object may induce an eccentricity in the inner binary via the Kozai mechanism \citep{1962AJ.....67R.579K}. \citet{2006A&A...450..681T} argues, for P$<$3d binaries, 96\% have a tertiary companion. The distortion of XY UMa's long-term optical light curve also indicates a third companion with a period of about 30 years \citep{1998A&A...340..415C,2001A&A...371..997P}. The relative strength between general relativity and the perturber \citep{2014ApJ...781L...5D,2007ApJ...669.1298F} is,\\
\begin{equation}
\rm \epsilon_{GR}=\frac{8G\rm M^2b_{per}^3}{c^2a^4\rm M_{per}} \approx 6.5\times 10^{-4}\ (Equation.2\ of\ Dong\ et\ al.\ 2014)
\end{equation}
if parameters like $\rm b_{per}$=10AU, a=1.5AU, M=2$\rm M_{\odot}$ and $\rm M_{per}$=0.23$\rm M_{\odot}$ \citep{1998A&A...340..415C} are adopted. Hence, in XY UMa, Kozai oscillation should not be suppressed by general relativity effect at least. Second, a non-early type binary in fact can have an elliptical orbit and short period. For example, \citet{2015IBVS.6138....1C} reported a P $\approx$ 0.86d binary KIC2835289 based on Kepler data. Its eccentricity is being detected by more observations. KIC2856960 \citep{2013ApJ...763...74L} is the other example in the literature, which is a two M-type binary with a 6.2 hr orbital period and small eccentricity $\approx$ 0.0064. Third, \citet{2007AJ....133.1977P} used a $sine$ curve to fit the radial velocities of XY UMa\footnotemark[8] directly, but did not assess significance. We re-fit the radial velocity data, assuming constant uncertainties, and find that we can rule out eccentricities larger than 0.01 from the data, but the data are not sensitive to eccentricities smaller than this value.

\section{Conclusions}

We used a serendipitously obtained observation to investigate the timing of flares
and size scales of coronal structures in a close binary system XY UMa.
The existence of two very energetic flares so close in time to each other is marginally consistent with
expected flare frequencies for active binary systems.  The lack of eclipses seen in the X-ray light curve
are consistent with most of the non-flaring X-ray emission being produced in a polar spot which is always visible.
Assuming there is a single flaring loop involved in the X-ray flare, analysis of the temperature and emission measure
reveal large length scales. Whether these are large enough to connect the magnetospheres of the two stars
and provide a trigger for flares at preferential orbital phases is suggestive, but not conclusive.
All three possibilities (tertiary companion, non-early type binary, sine curve to radial velocities)
for periastron occurring near primary eclipse
are speculative and marginally acceptable.
Even if the orbit of XY UMa is not so circular, whether a small eccentricity
can produce such an effect is not known.
Because there are only two flares, we hope future X-ray or radio monitoring can test whether flares have a significant accumulation near XY UMa's primary and secondary eclipse.
Such observations would shed light on how stellar coronal environments are shaped by
interactions with a companion.

\footnotetext[1]{http://cxc.harvard.edu/ciao/why/dither.html}
\footnotetext[2]{http://cxc.harvard.edu/proposer/POG/}
\footnotetext[3]{http://cxc.harvard.edu/ciao/dictionary/pileup.html}
\footnotetext[4]{http://cxc.harvard.edu/ciao/ahelp/pileup\_map.html}
\footnotetext[5]{http://cxc.harvard.edu/ciao/ahelp/wavdetect.html}
\footnotetext[6]{http://cxc.harvard.edu/ciao/ahelp/times.html; http://www.physics.sfasu.edu/astro/javascript/hjd.html}
\footnotetext[7]{http://cxc.harvard.edu/sherpa/ahelp/xswabs.html}
\footnotetext[8]{http://vizier.cfa.harvard.edu/viz-bin/VizieR-3?-source=J/AJ/133/1977/table1}

\normalem
\begin{acknowledgements}
We thank Paul Sell, Xinghua Dai, Xin Huang, Jingxiu Wang, Songhu Wang and Subo Dong for helpful discussions and improvements on the paper. We also thank two anonymous referees for detailed corrections which help to improve the quality of this paper. This research has made use of data obtained from the $\emph{Chandra}$ Data Archive, and software provided by the $\emph{Chandra}$ X-ray Center (CXC) in the application packages CIAO, ChIPS, and Sherpa.

The optical data in Figure.1 were made possible by very generous allocations of telescope time at Mount Laguna Observatory.

\end{acknowledgements}

{}

\end{document}